\documentclass[%
 aip,
 amsmath,amssymb,
 reprint,%
]{revtex4-1}

\usepackage{graphicx}
\usepackage{dcolumn}
\usepackage{bm}

\usepackage[utf8]{inputenc}
\usepackage[T1]{fontenc}
\usepackage{mathptmx}
\usepackage{etoolbox}
\usepackage{xcolor}
\definecolor{DK}{RGB}{255, 0, 0} 
\definecolor{DK_c}{RGB}{0, 0, 255}

\makeatletter
\def\@email#1#2{%
 \endgroup
 \patchcmd{\titleblock@produce}
  {\frontmatter@RRAPformat}
  {\frontmatter@RRAPformat{\produce@RRAP{*#1\href{mailto:#2}{#2}}}\frontmatter@RRAPformat}
  {}{}
}%
\makeatother

\begin{document}

\preprint{AIP/123-QED}

\title[]{Open whispering gallery modes resonators}
\author{D. R. Kazanov}
\affiliation{Ioffe Institute, Politekhnicheskaya 26, St. Petersburg, 194021, Russia} 

\author{A. M. Monakhov}%
\affiliation{Ioffe Institute, Politekhnicheskaya 26, St. Petersburg, 194021, Russia} 

\date{\today}

\begin{abstract}

There are some issues with traditional whispering gallery mode (WGM) resonators like poor light extraction and a dense mode spectrum. In this paper we introduce a solution to these limitations by proposing open WGM (OWGM) resonators that effectively reduce the mode density and enable directional radiation through a connected waveguide at the expense of some lowering in Q-factor. Numerical simulations of two-dimensional metallic and dielectric disk resonators with holes reveal a significant increase in intermode distance. The study also extends to three-dimensional dielectric OWGM resonators, demonstrating the formation of sparse spectra suitable for photonics applications. Additionally, the design of a cylindrical Bragg microresonator connected to a single-mode fiber via an optimized topology-based connector achieves near-unity transmission and efficient coupling. This approach enhances the development of new photonic devices, addressing the limitations of traditional high Q-factor WGM resonators and offering potential advancements in laser technology and optical communications, where compactness and manufacturability are prioritized over ultra-high Q-factors.

\end{abstract}

\maketitle

Open resonators are typically employed to mitigate the dense spectral characteristics inherent in closed resonators. The modes within these open resonators are confined either by internal caustics, as observed in barrel resonators \cite{Kleev1995}, or through reflections at the open ends of the resonators, such as in open-ended cylinders and Fabry-Pérot resonators. These systems often comprise a series of mirrors separated by an intervening open space. While barrel-shaped resonators exhibit a high quality factor, they also possess a significant mode density, which limits their effectiveness in spectral thinning \cite{Vaynshteyn}.

Another reason for using open resonators is the necessity of resonator coupling to an external optical circuit. This problem is especially relevant for resonators operating on whispering gallery modes (WGM).
Resonators operating on these modes are known to have very high Q-factor ($Q\sim 10^6$ and higher \cite{Q}). These resonators are quite compact, technological in production, and they are widely used as laser resonators, narrow-band filters in optical communications, optical sensors and so on \cite{Ward2011, Jiang2020, Vollmer2021}. Nevertheless, WGM resonator high Q-factor has a drawback, that is the issues with light extracting from a WGM resonator and the existence of a lot of high Q modes with the rather small intermode distances. For that reason, WGM lasers mostly operate in multimode regime \cite{Laser}. The dense mode spectra prevent the usage of WGM resonators for a single-photon light source due to the high probability of a multi-mode excitation of the resonator. This WGM light-extracting problem is usually solved by either making a small defect on the resonator surface \cite{defect} or waveguide tunnel coupling \cite{Cai2020}. Both of these methods  take apart the high mode density problem

There are some areas of possible WGM resonators application where one is interested in not a high Q-factor, but their compactness and manufacturability. In this case, it is often necessary to thin out the mode spectrum of the resonator. This problem has been solved for the common lasers using open Fabry-Per\'ot resonators where light is confined in one direction and can be freely emitted in others. In this paper, we suggest the WGM resonator construction that have rather low mode density and Q-factor acceptable for laser and nanophotonic applications.

\begin{figure}[t]
    \centering
    \includegraphics[width=0.99\columnwidth]{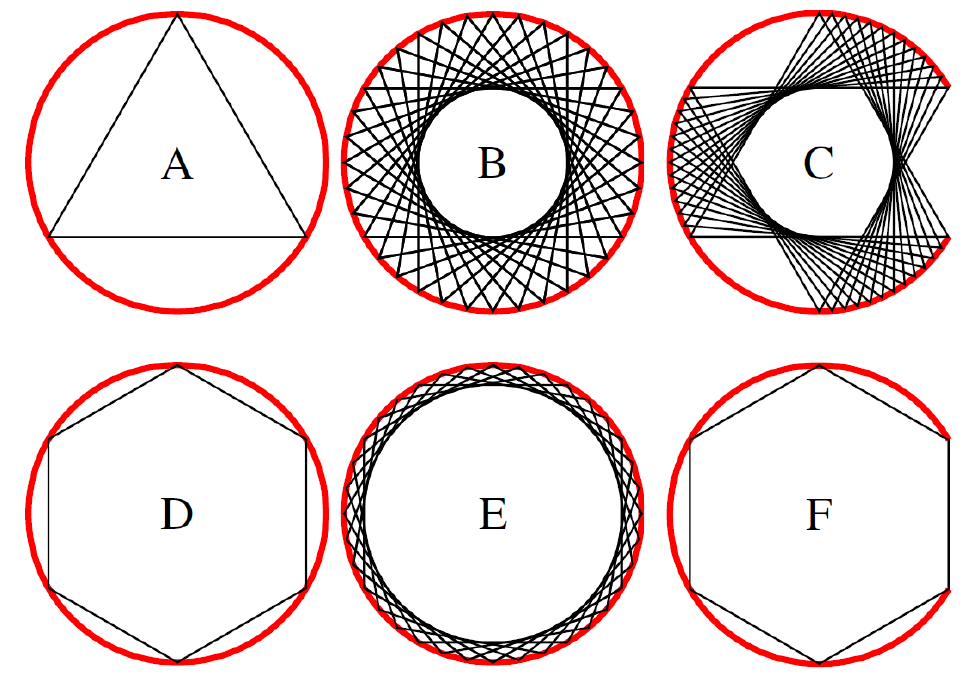}
    \caption{Example of the simplest rays and their subset of normal congruences for closed and open circular resonator. Top row: ray (A) and subset of the normal congruence of the ray in a closed (B) and open (C) resonator. Bottom row: ray (D) and subset of the normal congruence in a closed (E) resonator, while (F) shows that it cannot create a normal congruence in the open resonator. Corresponding mode exists in the closed resonator and disappear in the open one.}
    \label{fig:WGM}
\end{figure}

\begin{figure*}[t]
    \centering
    \includegraphics[width=1.99\columnwidth]{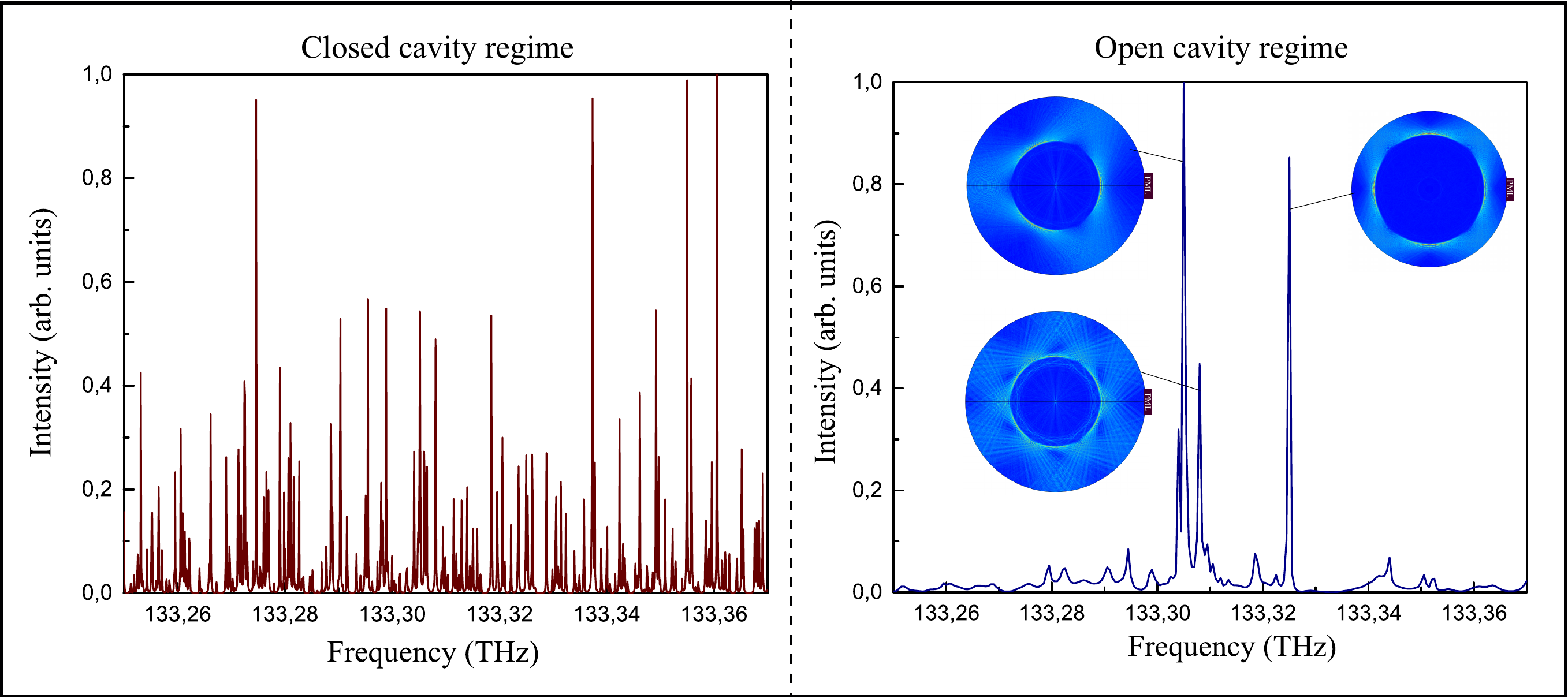}
    \caption{Spectra of larger radius 2D cylindrical resonator that has closed boundaries (left) and a hole on the side (right).}
    \label{fig:2DPL}
\end{figure*}
For the sake of simplicity, we will first focus on the two-dimensional case. Initially, we aim to examine a two-dimensional circular metallic resonator from the perspective of quasi-geometrical optics.  It is established \cite{babich} that in the short-wave approximation, modes within a resonator emerge from a set of closed rays having the same optical length and reflecting from the walls of resonator.  Mathematicians refer to this set as normal congruence (NC). Such rays are shown in Fig. \ref{fig:WGM}A and B. The subset of NC originated from these rays in the close circular resonator are shown in Fig. \ref{fig:WGM}B and E. Clearly there exist more intricate NCs where a ray completes multiple revolutions around the circle before returning to its statring point. All these NCs contribute to the dense eigenfrequency spectrum of the resonator.

When a hole is created in the wall of the resonator (Fig. \ref{fig:WGM}C and F), only the rays that do not enter the hole will persist. For example, mode B will remain intact in the open resonator (see Fig. \ref{fig:WGM}C) while mode E disappeared because the ray in Fig. \ref{fig:WGM}D is non-able to form NC in the open resonator. Summarizing, only the modes resulting from the rotations of the digon, triangle, and square are retained. Modes associated with polygons that have more than five sides or that complete multiple rotations around the circle disappear, as they cannot form a NC (Fig. \ref{fig:WGM}F). This phenomenon allows for a reduction in the spectrum of the WGM resonator, thereby increasing the distance between modes.

Introducing a ''hole'' in the resonator facilitates the control of directional radiation output, as this function can be effectively fulfilled by a matched waveguide that is coupled to the resonator. This configuration allows for a more controlled and efficient extraction of energy from the resonator. It is important to note that the Q-factor of an open whispering gallery mode (OWGM) resonator will inherently be lower than that of a conventional closed resonator. However, in certain applications, excessively high Q-factors may not only be unnecessary but could also be detrimental to performance. For instance, in scenarios where rapid signal processing or broad bandwidth is required, a lower Q-factor can enhance the system's responsiveness and reduce sensitivity to environmental perturbations.

The simple short-wave approximation shows that the eigenfrequency of the mode shown in Fig. \ref{fig:WGM}B should not differ strictly from that in Fig. \ref{fig:WGM}C. Indeed, in this approximation the phase incursion on a closed ray should be equal to the integer number of $2\pi$ including additional phases appears from reflections and caustic touches. Rays in Fig. \ref{fig:WGM}B and C have the same length, number of reflections and caustic touchpoints, so they have the same eigenfrequency in this approximation. The eigenfrequency difference between modes in closed and open resonators arises due to the escape of radiation from the open resonator, leading to the appearance of an imaginary part of the frequency and to the shift of the real part of it. This shift is small if the resonator has high enough Q-factor. 

The analytical solution for the field within an OWGM resonator, necessary for the exact eigenfrequency calculation, can hardly be obtained analytically from Maxwell's equations. While it is possible to obtain such a solution for a perfect metal cylinder with a cut in the wall using the Wiener-Hopf technique outlined in reference \cite{Vaynshteyn}, the resulting expression is quite complex. Ultimately, numerical calculations are still required to visualize the electromagnetic field distribution. Therefore, we validate the aforementioned qualitative reasoning through numerical simulations.

In the Fig. \ref{fig:2DPL} the excitation spectra for the closed and open resonator of the perfect metal 100 $\mu$m in diameter are shown. The hole has been emulated by a perfectly matched layer that absorbs the incident radiation almost completely. This boundary condition makes it possible to avoid the issue with the divergence of the field for the Sommerfeld radiation condition (see \cite{klimov} and bibliography therein). The resonators were excited by a dipole, placed at the same point. The map of the electric field module in the  OWGM  resonator is shown in inserts on the right plot. The small rectangle on the right side of resonators is the absorption layer (PML). The OWGM modes have complex-valued eigenfrequencies $\nu=\nu'-i\nu'',$ and the mode Q-factor can be estimated as $Q=\frac{\nu'}{2\nu''}.$ Our calculations show that the Q-factor of the modes is about $Q\sim 2\cdot10^5.$
\begin{figure}[t]
 \includegraphics[width=0.99\columnwidth]{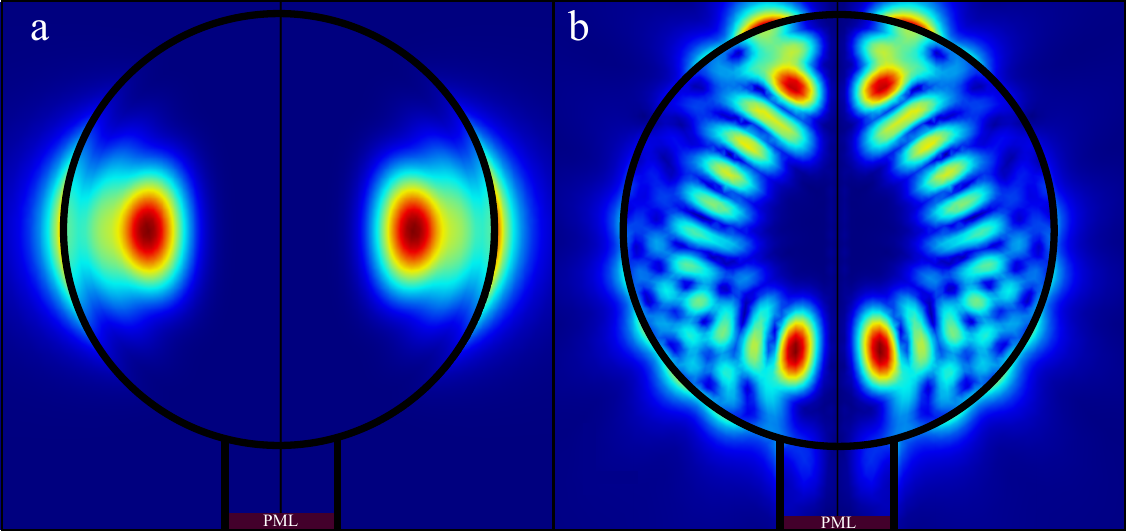}
 \caption{OWGM in  3D sphere  10 $\mu$m in diameter. (a) shows high-Q mode originates from the common WGM in the spherical resonator, obtained by rotation of the Fig. \ref{fig:WGM}A around the polar axis orthogonal to the picture. (b) Distribution of electromagnetic field module corresponds to $m=3$ according to \eqref{Ora} and originated from 2D mode shown in Fig. \ref{fig:WGM}B rotated around the picture symmetry axis.}
 \label{fig:OWGDiel}
 \end{figure}
 
Moving on to a more realistic scenario involving three-dimensional dielectric resonators, we should point out that the OWGM in three-dimensional resonators are more complex due to the presence of an extra degree of freedom. We first consider OWGM spherical resonators, which can be readily produced by melting the end of a dielectric optical fiber. The fiber attached to the sphere plays a role of a hole in the resonator. Some modes of an open spherical resonator are described in literature (see, for instance, \cite{Vaynshteyn}). A notable characteristic of these modes is that they are confined by the internal caustic and exhibit high quality factors, as the electromagnetic radiation diminishes rapidly beyond the caustic.
 
Looking at Fig. \ref{fig:WGM}B,C one can easily guess the existence of two types of modes in such a resonator. It should be noted that these two types do not exhaust the list of modes possible in this resonator.  Recalling the quasi-geometrical approach, one can say that in the OWGM spherical resonator the mode is formed by a set of rays similar to those shown in Fig. \ref{fig:WGM}B,C rotated around some axis tilted so that the rays do not fall into the hole.  A bit more rigorously, one can say that the Debay potential for the modes of a spherical dielectric resonator in the spherical coordinates have the form \cite{Oraevsky}:
\begin{equation}\label{Ora}
	U(r,\theta,\varphi)=\frac{C}{\sqrt{kr}}P_n^m(\cos\theta)Z_{n+\frac{1}{2}}(kr)e^{\pm i m \varphi},
\end{equation}
where $C$ is a normalization constant, $k$ -- wave vector, $P_n^m$ -- associated Legendre polynomial and $Z$ is a Bessel function. The associated Legendre polynomial $P_n^m(x)$  decreases rapidly when $x\to\pm 1$ for large $n$, so if one drill a small enough hole around the rotation axis, the resonator mode will hardly notice it. It means that the WGM and OWGM can be almost the same for large enough $n$ when the rotation axis is close to the polar one. The coupling between the resonator and the waveguide, playing the role of a hole, can be controlled by an appropriate choice of the mode index $n$ and the waveguide diameter. In Fig.~\ref{fig:OWGDiel} the numerical calculation of the electric field distribution in dielectric sphere 10 $\mu$m in diameter with refractive index $n=3.4$ is shown. The figure (a) corresponds to the mode arising from the NC obtained by rotation of Fig.~\ref{fig:WGM}B around the axis slightly inclined to perpendicular to the plane of the drawing. This is the common mode of the open spherical resonator described in \cite{Vaynshteyn}. The NC for the figure (b) obtained by rotation of Fig.~\ref{fig:WGM}C around the picture symmetry axis. To the best of our knowledge, this mode is not described in literature.

We would like to point out that OWGM spherical resonator has a high Q-factor but the dense mode spectrum. The high Q-factor of the mode is due to the fact that the mode in it is protected by an external caustic formed by the tangent surface to the NC forming the mode. A dense mode spectrum arises because a significant part of the closed resonator modes is present in the open one for the reasons mentioned above. In the short-wave approximation, the mode density can be estimated as \cite{Vaynshteyn}
$$\Delta N=\frac{S}{\pi c^2}\omega\Delta\omega,$$
where $N$ is the number of modes in the frequency interval $\Delta\omega$ around $\omega$ and $S$ is the diametrical cross-sectional area of the resonator.  Thus, an open spherical resonator is not the most suitable device for spectrum thinning. 

\begin{figure}[t]
    \centering
    \includegraphics[width=0.99\columnwidth]{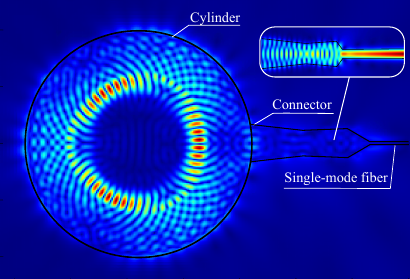}
    \caption{OWGM in 3-D dielectric sphere 10 $\mu$m in diameter. Figure (a) shows common mode that originates from the WGM in the spherical resonator, obtained by rotation of the Fig. 1B around the polar axis orthogonal to the picture. (b) The mode corresponds to m = 3 according to \cite{Vaynshteyn} and originated from 2-D mode shown in Fig. 1C rotated around the picture symmetry axis.}
    \label{fig:FOM}
\end{figure}

A more promising device for this application can be a short cylindrical resonator but with an open side. Here, "short" refers to a height that is comparable to the operating wavelength, while its diameter significantly exceeds the wavelength. A widely utilized cylindrical resonator is the three-dimensional Bragg resonator, which has shown considerable efficiency as a component for single-photon sources \cite{Solomon2001, Senellart2017}. However, its design requires substantial modifications, as conventional implementations predominantly depend on lower angular modes, such as HE$_{11}$. Thus, we carried out calculations on a three-dimensional structure that supports WGMs for a specific wavelength of 925 nm, as illustrated in Fig. \ref{fig:FOM}. This structure comprises three components: a cylindrical Bragg microresonator with a large radius, a single-mode fiber serving as an aperture, and an optimized connector linking them. In traditional GaAs/AlAs Bragg microresonators, as discussed in \cite{Galimov2021, Rakhlin2023}, the primary wavelength $\lambda_0$ is selected for emission, and the layer heights are determined based on the Bragg condition: $\lambda_0/4n_j$ for the layers and $\lambda_0/n_j$ for the active layer, where $n_j$ denotes the refractive index of the respective layer. Consequently, such a structure supports the first optical waveguide mode at a wavelength close to $\lambda_0$. However, our focus is on WGM within an open cavity featuring a side fiber. A significant challenge arises as resonance frequencies cannot be derived analytically due to the inseparability of variables in Maxwell's equations. Thus, in the initial approximation, we can use the expression $(2\pi/\lambda_0)^2 \sim k_{||}^2+(2\pi/\lambda_1)^2$, where $\lambda_1$ is a wavelength greater than $\lambda_0$ and $k_{||}$ s the wavevector in the plane of the layer associated with the WGM mode order. The choice of $\lambda_1$ only affects the dimensions of all layers in the $z$-direction. By optimizing both $\lambda_1$ and the radius of a cylinder, we can achieve the desired mode at wavelength $\lambda_0$.

Nonetheless, coupling a wave into a single-mode fiber is not a simple task due to significant reflectivity at the interface between the large cylinder and the fiber. To address this issue, we developed a connector designed to minimize reflection and effectively "open" our cavity. To create a connector with high transmission efficiency to the fiber at a specific wavelength, we employed an adjoint method of topology optimization \cite{Yablonovich2013}. This method leverages the reciprocity of Green's function $G(x,x_0)=G(x_0,x)^T$, where $G(x,x_0)$ represents Maxwell's Green’s function links the electric field at point $x_0$ and the induced polarization $p^{\rm ind}$ at a point $x$ within a small volume. This approach allows us to optimize the figure-of-merit (in our case, transmission approaching unity) using only two calculations: forward and adjoint. The gradient of the figure-of-merit is computed to facilitate geometric adjustments that correlate directly with this gradient. This gradient descent technique can be iteratively applied to converge on an optimal solution. The initial geometry was selected based on a tapered design, transitioning from a small disk segment to a single-mode fiber, with a length of several micrometers. The optimization algorithm modified the external boundaries of the connector to facilitate wave propagation from the left edge to the right edge with minimal losses and reflections. The inset of Figure \ref{fig:FOM} illustrates the calculation of the electromagnetic field over an extended time period to demonstrate the leakage of the field from the system. Figure \ref{fig:FOM} displays the electromagnetic field distribution module within the optimized connector geometry, demonstrating field transfer efficiency close to unity from the left side of the structure to the single-mode fiber on the right. Achieving near-unity transmission in practical applications presents significant challenges, as the fabricated structure must closely adhere to the precise specifications established during the modeling phase. Any deviations in the manufacturing process or material properties can lead to considerable discrepancies in performance, thereby undermining the intended optical characteristics.

All calculations were carried out using Ansys Lumerical software, while post-processing of optimization data was carried out using Python and Lymopt libraries. We have constructed a large simulation box with perfectly matched layers to minimize non-physical reflections from its boundaries. Within this box, we established a Bragg resonator oriented in the $z$-direction, incorporating an active $\lambda$-cavity layer. In the $x$ and $y$ directions, we selected specific geometries for the cylinder, connector, and single-mode fiber. Light sources were modeled as dipole sources within the active layer and positioned according to pre-calculated eigenvectors at points of maximum electric field for the chosen mode. Photoluminescence (PL) intensity was derived by integrating the modulus of the squared electric field within the active layer.

\begin{figure}[t]
    \centering
    \includegraphics[width=0.99\columnwidth]{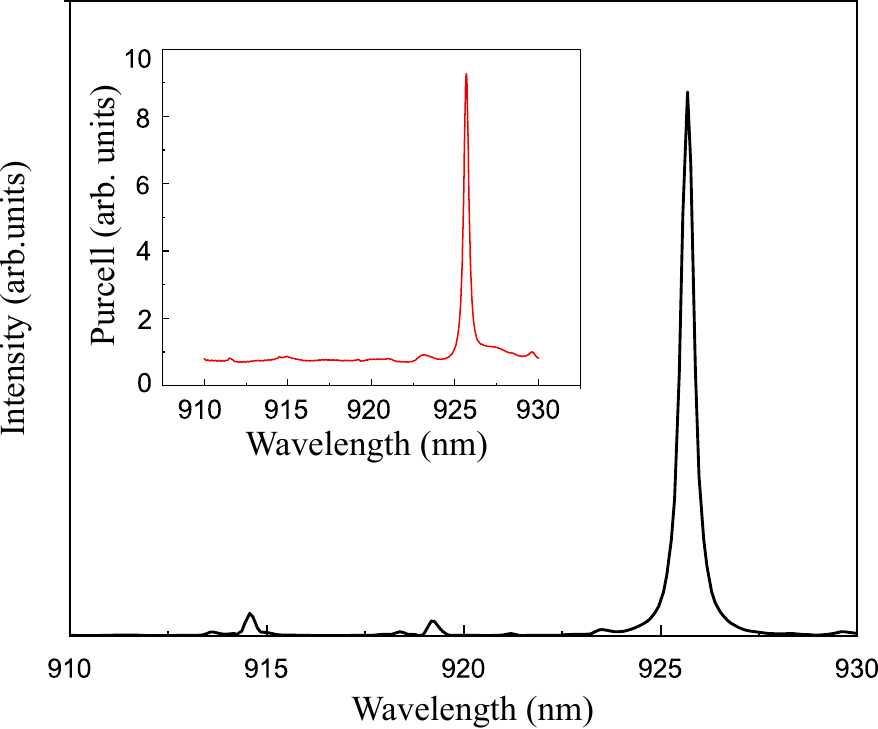}
    \caption{PL spectrum of a studied structure consisted of cylinder, connector and single-mode fiber. Inset shows a Purcell factor of dipole that is located in the maximum of electromagnetic field.}
    \label{fig:PL_3D}
\end{figure}

Ultimately, this connector effectively "opens" our cylindrical Bragg cavity, resulting in a sparse PL spectrum that can be characterized by only a few modes. Figure \ref{fig:PL_3D} presents an example of PL spectra for the analyzed resonator, revealing several modes with substantial intermode spacing exceeding 5 nm. Furthermore, we exclusively excite those modes whose maximum electric field distributions align with the location of the light source in the layer, facilitated by the Purcell effect \cite{Gerard1998}.

In this paper, we have investigated specific modes within open whispering gallery mode (OWGM) resonators, demonstrating their capability to selectively thin out the mode spectrum at the expense of some reduction in the Q-factor of the resonator. Such a device can be a WGM resonator with a waveguide of a sufficiently large size connected to it. This approach addresses the inherent limitations of conventional high-Q WGM resonators, particularly in applications where compactness and manufacturability are paramount, such as in laser technology and optical communications. Furthermore, we have introduced a numerical model for a single-photon source Bragg resonator optimized for coupling with single-mode fibers via a connector, employing the adjoint method of topology optimization. Our results confirm that single-mode fibers can serve as effective apertures for resonators at designated wavelengths, paving the way for enhanced performance in photonic applications.

\begin{acknowledgments}
This work was supported by Rosatom  in the framework of the Roadmap  for Quantum computing (Contract No. 868-1.3-15/15-2021 dated 05.10.2021 and No. P 2152 dated 11.19.2021). D. R. K. also acknowledges the grant \#SP5068.2022.5.
\end{acknowledgments}

\end{document}